\documentclass{ws-p8-50x6-00}

\begin{document}

\title{A new interpretation of the QCD phase transition and of strangeness
as QGP signature}

\author{Sonja Kabana}

\address{University of Bern, Sidlerstrasse 5, 3012 Bern, Switzerland
\\
E-mail: sonja.kabana@cern.ch}

\maketitle

\abstracts{
We address the question of how to identify the QCD phase
transition using measured light (u,d,s-structured)
 hadrons, without invoking comparison
to the QCD $\epsilon_c$ predictions, and extract $\epsilon_c$ from the data.
We analyse several particle and nuclear collisions and
extract their chemical freeze-out temperature $T$
 at zero baryochemical potential ($\mu_B$).
We  find at $\mu_B=0$
 a universal rise and saturation of both the $T$ 
and of the strangeness suppression factor $\lambda_s$
(=$\frac { 2\overline{s} } { \overline{u} + \overline{d} } $)
 with increasing initial energy density ($\epsilon_i$).
The onset of saturation
of both $T$ and $\lambda_s$, is interpreted 
as due to the event of the QCD phase transition.
The critical energy density is estimated
to be $\epsilon_c$ $\sim$ 1 +0.3 -0.5 GeV/fm$^3$,
corresponding approximately to a $\sqrt{s}$ of $\sim$ 8.8  GeV for 
central Pb+Pb collisions. 
Concerning the role of strangeness, we identify trivial and non-trivial
sources of strangeness enhancement:
The peak of $\lambda_s$  in  Pb+Pb collisions at $\sqrt{s}$=8.8 GeV
and other phenomena of 'strangeness enhancement' defined with respect to
 p+p data,
are trivially traced back to the different baryochemical potentials
and $\epsilon_i$ of the compared systems.
A non trivial redefined '$\lambda_s$ enhancement' is however also present.
The netbaryonfree $\lambda_s$
limit is estimated to be approximately reached in Au+Au
collisions at the LHC.
}

\section{Introduction}
It has been repeateadly demonstrated in the literature 
\cite{pbm,redlich_qm2001,mapping} 
that in many cases the final state of nuclear and particle collisions
is  compatible with the hypothesis of an equilibrated source.
In the following the discussion will concern only colliding systems
for which this finding holds (measured by the $\chi^2$ of thermal
model fits).
Much work is presently concentrated on identifying
the QCD phase transition at a certain $\sqrt{s}$,
separating the colliding systems which go through the phase transition
from those which dont.
 This defines the problem that we address. 
One way to approach this problem is to compare the estimated
thermodynamic parameters temperature ($T$) and baryon chemical
potential ($\mu_B$) at the chemical freeze-out
extracted by thermal models from the data, 
to the expected critical ($T_c$, $\mu_B(c)$) values of QCD (e.g. \cite{pbm}).
However, the theoretical expectations
 for the critical temperature $T_c$ are uncertain.
It is thus obviously interesting to try to extract informations
on the critical parameters from the data, without any use of the
QCD predictions.
In this spirit, we approach this problem  in the present study
in a different way, namely
  by extrapolating the data to equivalent systems with zero $\mu_B$
and studying their $T$ \cite{sqm2000}.
To see why this is interesting, we follow a line of arguments :
If all measured colliding systems above a certain $\sqrt{s}$
are heated enough to go through the phase
transition and back, they would appear to have the same
$T$ (if they don't freeze-out in a considerably different
way) which we note as 'limiting' hadronic temperature $T_{lim}$.
 The colliding systems which do not reach the $T_c$ at any
time because of their small $\sqrt{s}$,
 will exhibit a final $T$  smaller than $T_{lim}$. 
Therefore, in order to separate the colliding systems
which went through the transition from the ones which did not,
 one can investigate the $T$ of all colliding systems at conditions
of $\mu_B$=0, as a function of $\sqrt{s}$ or $dN/dy$ at midrapidity,
 and search for an increase of $T$ followed by a saturation starting
at the $\sqrt{s}$ where the $T_c$ is reached.
This would work if we use always the same projectile and
target combination.
Otherwise, one should
correct for the fact that different projectile/target
combinations at the same $\sqrt{s}$, reach different initial energy densities.
For this reason, we will investigate the $T$ as a function
of the initial energy density ($\epsilon_i$) reached at each collision after
1 fm/c based on the Bjorken estimate \cite{bjorken} and also using
other Ans\"atze especially at low $\sqrt{s}$ \cite{mapping}. 
For a discussion on the uncertainty on $\epsilon_i$ 
see \cite{charm,mapping,border}.
The same conclusions can however be reached by the present analysis, 
while looking only at one projectile/target combination A+A,
namely with $A \sim 200$ (that is at Pb+Pb and Au+Au central collisions),
 as a function of e.g. $\sqrt{s}$ instead of $\epsilon_i$.

\section{Energy density dependence of temperature and $\lambda_s$}

\begin{figure}[ht]
\vspace*{-0.5cm}
\begin{center}
\mbox{\epsfig{file=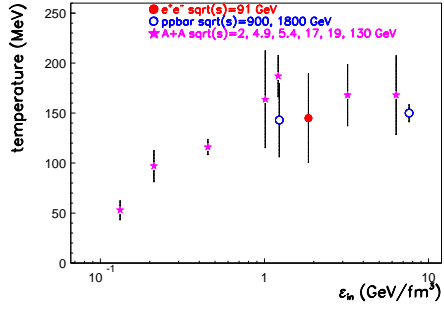,width=58mm}}
\end{center}
\caption{
The temperature extrapolated to zero fugacities along an isentropic path,
as a function of
the initial energy density for several nucleus+nucleus, hadron+hadron and
lepton+lepton collisions.
We demand for the fits confidence level $>$ 10\%.
}
\vspace*{-0.5cm}
\label{t_vs_e}
\end{figure}

\noindent
\underline{ \bf 1. $T$:}
We compare the ratios of experimentally measured
hadron yields in nuclear collisions with the prediction of
a thermal model of non interacting free hadron resonances
(for details see \cite{mapping}).
We  extract the thermodynamic parameters describing best
 the particle source: temperature, baryochemical potential ($\mu_B$)
and strangeness chemical potential ($\mu_S$), imposing
exact strangeness conservation.
We then extrapolate all thermodynamic states with parameters
($T$, $\mu_B$, $\mu_s$) to equivalent states
at zero chemical potentials ($T(\mu_B=0)$, 0, 0) along isentropic paths.
As demonstrated in figure \ref{t_vs_e} the resulting temperature at $\mu_B=0$
rises and saturates above $\epsilon_i \sim$ 1 $GeV/fm^3$.
We interpret the onset of saturation of $T$ as due to 
the reach of an initial temperature greater or equal to $T_c$.
We therefore extract from the onset of saturation in fig. \ref{t_vs_e}
 the 'critical' $\sqrt{s}$ of $\sim$ 
 8.8 GeV/nucleon pair respectively the critical
 $\epsilon_c$ of $\sim$ 1 +0.3 -0.5 GeV/fm$^3$ of the QCD phase transition, 
 independently of the QCD predictions.
\noindent
An important assumption needed to support this interpretation is
that the cooling until freeze-out of the particle source,
is not significantly (as compared to the errors)
dependent on the $\sqrt{s}$, in the range
between $\sqrt{s}$=2 and 9 GeV for A+A collisions.
\footnote{Note that a
 similar behaviour (rise and saturation) was found in the $\epsilon_i$ 
dependence of the kaon number density $\rho_k$ \cite{charm,centr_na52},
which may be related to the behaviour seen in fig. \ref{t_vs_e}.}
A discussion of $T$ rise and saturation can be found in \cite{horst},
but not for the $\mu_B$=0 case.
However, it is only when investigating systems with the same $\mu_B$,
where this behaviour is expected to be exact.
The importance of using a common $\mu_B$ \cite{sqm2000,mapping,border},
becomes more apparent when investigating strangeness. 
 This is the next topic.
 
\begin{figure}[ht]
\vspace*{-0.5cm}
\begin{center}
\mbox{\epsfig{file=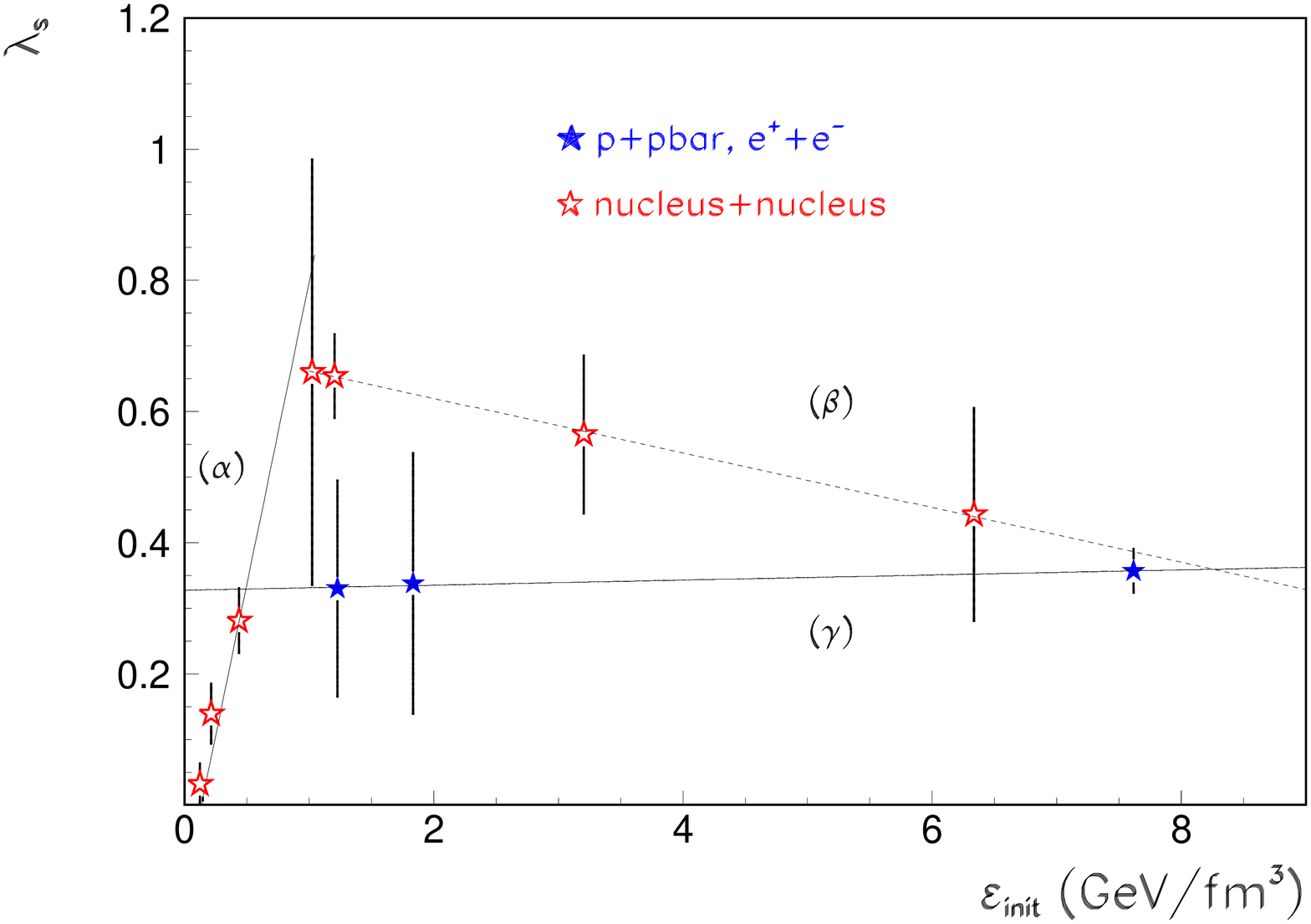,width=58mm}}
\mbox{\epsfig{file=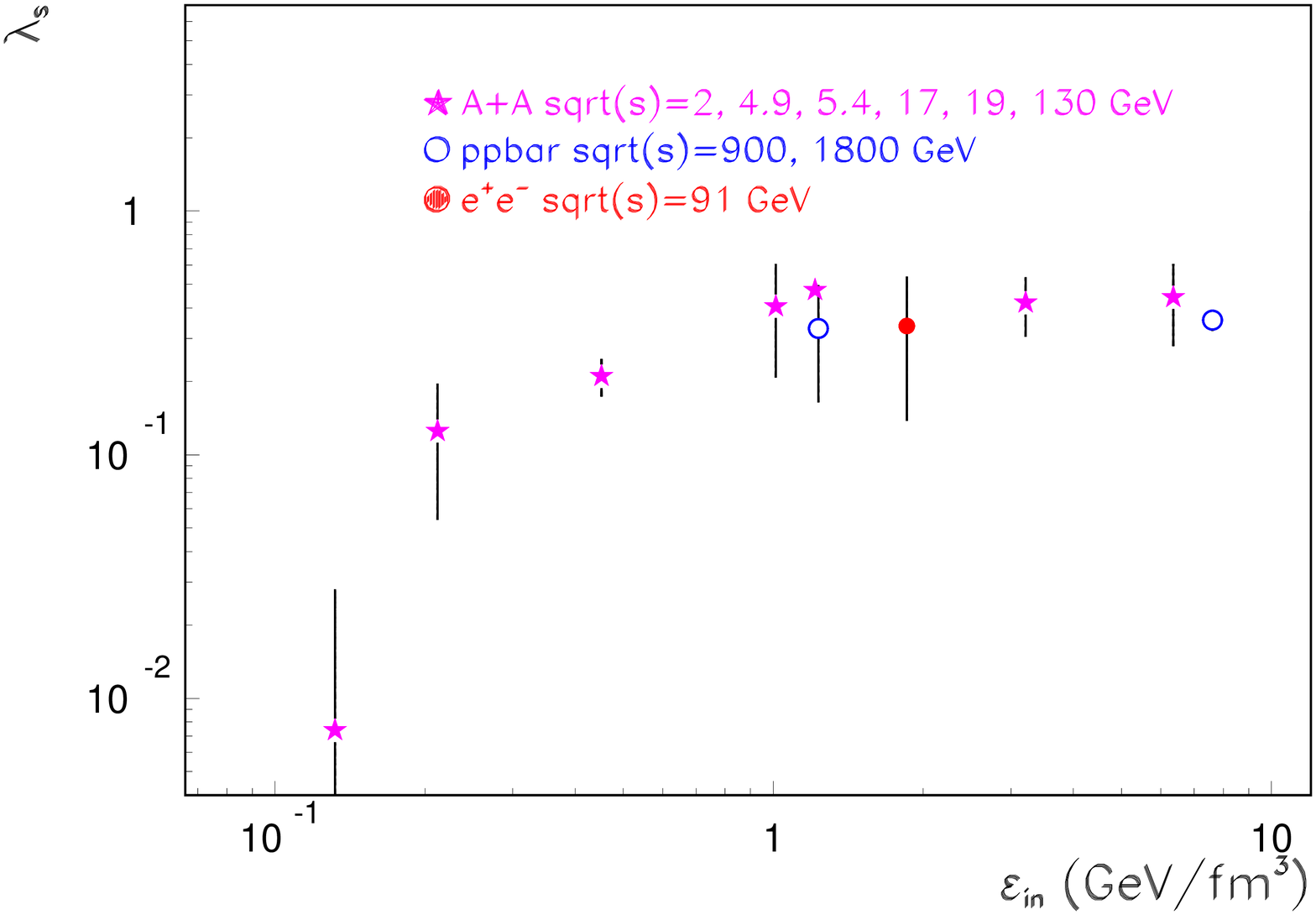,width=58mm}}
\end{center}
\caption{
The $\lambda_s$ factor as a function of
the initial energy density for several nucleus+nucleus, hadron+hadron and
lepton+lepton collisions. Left for non zero $\mu_B$, right for
zero $\mu_B$.
Left, the lines $\alpha$ and $\beta$ show $\lambda_s$ at
nonzero $\mu_B$, while the line $\gamma$
show $\lambda_s$ at zero $\mu_B$.
We demand for the fits confidence level $>$ 10\%.
}
\label{ls_vs_e}
\vspace*{-0.5cm}
\end{figure}

\noindent
\underline{\bf 2. $\lambda_s$:}
The so called 'strangeness suppression factor' 
$
\lambda_s \ = \
\frac{ (2 \overline{s}) }
{ (\overline{u} + \overline{d}) }
$
at nonzero $\mu_B$,
rises with $\epsilon_i$ up to 1 GeV/fm$^3$ (line (a) in  fig. \ref{ls_vs_e},
 left) and then decreases (line (b) in  fig. \ref{ls_vs_e}, left).
After extrapolating to $\mu_B=0$, $\lambda_s$ rises and saturates
universally above $\epsilon_i$ $\sim$  1 GeV/fm$^3$ (fig. \ref{ls_vs_e},
right).
Therefore the peak of $\lambda_s$ at the 40 A GeV Pb+Pb
(point at $\epsilon_i$ = 1 GeV/fm$^3$ in fig. \ref{ls_vs_e}, left) is due to
the high $\mu_B$ reached there. 
Furthermore, the so called 'strangeness enhancement' defined
usually 
as double ratio of strange/pion ratio in A+A over p+p collisions at the
same $\sqrt{s}$,
 can be traced back to the different $\mu_B$ and
$\epsilon_i$ of those reactions. In particular this explains why the
so defined strangeness enhancement increases with decreasing $\sqrt{s}$
(fig. 7 in \cite{redlich_qm2001}),
 since this corresponds to increasing $\mu_B$.
If one eliminates the bias due to different $\mu_B$
 by considering only $\lambda_s$ at $\mu_B=0$,
the non-trivial -with respect to the phase transition-
 'strangeness enhancement' is revealed (fig. \ref{ls_vs_e}, right)
as a consequence of the behaviour of $T$ seen in fig. \ref{t_vs_e}.
We therefore redefine the 
'strangeness enhancement', as an enhancement of
$\lambda_s$ in all thermalised systems above $\epsilon_c$, 
as compared to all thermalised systems
with $\epsilon_i < \epsilon_c$, at the same $\mu_B$.
\footnote{Discussions of the QCD phase transition appearing possibly 
between AGS and SPS energy 
can be found e.g. in \cite{dieter}, however not at a common $\mu_B$.}
We also 
analysed for this conference
 ratios from $e^+ e^-$ collisions at $\sqrt{s}$=3.6 GeV \cite{dasp}
and found a T of 124 +15 -20 MeV, and $\lambda_s$=0.25 +0.06 -0.09
with $\chi^2/DOF$=4 10$^{-3}$/1.
This T is below $T_{lim}$, however the $\epsilon_i$ is difficult
to extract due to lack of data. 
Extrapolating the A+A data in fig. \ref{ls_vs_e}, left,
 to higher $\epsilon_i$, we find that the approximately
netbaryonfree limit of $\lambda_s$ in Au+Au collisions,
is expected to be reached at LHC energies \cite{border}.

\vspace*{-0.2cm}

\section*{Conclusions}

The problem we address is how to identify the onset of the
QCD phase transition using measured light (u,d,s)-flavoured hadrons, 
and separate the colliding systems
which go through the QCD phase transition from the ones which dont,
without any use of the QCD predictions, as well as the role of strangeness.
We achieve this by estimating for the first time 
\cite{sqm2000,mapping,border} the equivalent temperature
at zero baryochemical potential
 of the thermodynamic states describing several
measured particle and nuclear collisions.
We find at zero chemical potential ($\mu_B$=0)
a universal rise and saturation of both the $T$ 
and of the strangeness suppression factor $\lambda_s$
(=$\frac { 2\overline{s} } { \overline{u} + \overline{d} } $)
 with increasing initial energy density ($\epsilon_i$)
(fig. \ref{t_vs_e} and fig. \ref{ls_vs_e}, right).
The onset of saturation
of both $T$ and $\lambda_s$ at $\mu_B=0$, allows to discriminate systems
which go through the QCD phase transition from those
which dont. The critical energy density is thus estimated
at the onset of saturation 
to be $\epsilon_c$ $\sim$ 1 +0.3 -0.5 GeV/fm$^3$,
corresponding approximately to a $\sqrt{s}$ of $\sim$ 8.8  GeV for central
 Pb+Pb collisions. 
Further, 
we identify trivial and non-trivial (with respect to the phase transition)
sources of 'strangeness enhancement':
e.g. the maximum of $\lambda_s$  at the 40 A GeV Pb+Pb collisions \cite{border},
and other phenomena of 'strangeness enhancement',
are 'trivially' traced back to the different baryochemical potentials
and $\epsilon_i$ of the compared systems (compare fig. \ref{ls_vs_e}
right with $\mu_B$=0 and left with nonzero $\mu_B$).
This explaines why e.g. the strange/pion double ratios in A+A over
 p+p collisions both at the same $\sqrt{s}$, increase with
decreasing $\sqrt{s}$, because $\mu_B$ increases and $\epsilon_i$
differs for A+A and p+p collisions at the same $\sqrt{s}$.
It could explain also an enhancement of $\lambda_s$ in p+A over p+p collisions.
We redefine the non-trivial 'strangeness enhancement',
as the increase of $\lambda_s$ in all (thermalised) systems
which reached $\epsilon_c$ as compared to all
(thermalised) systems which did not, at the same $\mu_B$.
It is a consequence of the $\epsilon_i$ dependence of $T$.
The netbaryonfree $\lambda_s$
limit is estimated to be approximately reached in Au+Au
collisions at the LHC \cite{border}.
In conclusion, in contrast to external non-equilibrium signatures like 
charmonia suppression or jet quenching which are important but
may be overcritical, 
hadrons with $u, d, s, \overline{u}, \overline{d}, \overline{s}$ quarks
play a special role as  equilibrium signatures and part of the plasma itself,
in allowing to extract the critical parameters of the QCD phase transition.

\section*{Acknowledgments}
We thank the Schweizerischer Nationalfonds for their support,
as well as K. Pretzl and P. Minkowski for fruitfull discussions.
We also thank the organisers of ISMD2001 for creating an
 open and scientifically fruitfull atmosphere.

\end{document}